\documentclass[11pt]{article}
\usepackage[margin=1in]{geometry}
\usepackage{amsmath,amssymb}
\usepackage{graphicx}
\usepackage{booktabs}
\usepackage{listings}
\usepackage{xcolor}
\usepackage{url}
\usepackage{hyperref}
\usepackage{algorithm}
\usepackage{algorithmic}
\usepackage{multirow}
\usepackage{titlesec}
\usepackage{abstract}
\usepackage{tikz}
\usetikzlibrary{arrows.meta, patterns, fit, backgrounds}

\titleformat{\section}{\normalfont\large\bfseries}{\thesection}{1em}{}
\titleformat{\subsection}{\normalfont\normalsize\bfseries}{\thesubsection}{1em}{}

\title{\textbf{The Art of Closed-Formula Defaults:}\\
\large Search-Free Code Generation for Tensor Operators}

\author{
  Paolo D'Alberto\footnote{Advanced Micro Devices, Inc. This work was developed in active collaboration with Claude (Anthropic).}\\
  \texttt{paolo.dalberto@amd.com}
  \and
  Ashish Sirasao\footnote{Advanced Micro Devices, Inc.}\\
  \texttt{ashish.sirasao@amd.com}
}
\date{}

\begin{document}
\maketitle

\begin{abstract}
Agentic search and automated optimization of GPU kernels are powerful
tools for large language model inference.
Their effectiveness, however, depends not on the sophistication of the
search itself, but on the clarity of the optimization problem being
solved.
We provide an application-first approach that drives a hierarchical
code generation tool from operator specification down to GPU
instructions, and show that a clearly defined computational model
makes the optimization problem tractable.

For four core tensor operators (general matrix multiply, 2D
convolution, softmax, and fused attention), a simple-to-describe,
CU-centric model determines all tile sizes analytically, without search:
local data share (LDS) capacity fixes the CU tile, the
vector general-purpose register (VGPR) budget fixes the warp tile,
and the number of compute units (CUs) bounds the minimum parallelism.
When the problem is clearly stated, the solution is a closed formula.
The resulting \emph{closed-formula defaults} are correct by construction
and derived once per hardware target; changing the hardware descriptor
re-runs the same derivation for a different compute DNA (CDNA) GPU
without any re-derivation by an expert.
These defaults stand on their own or serve as warm starts for any
subsequent automated search.

We validate on an AMD CDNA GPU.
General matrix multiply achieves 12.3~TFLOPS (53\% of the 23.1~TFLOPS
MFMA peak) using FlyDSL as code generation backend.
Fused attention with matrix fused multiply-add (MFMA) instructions for
both internal matrix multiplications achieves $1.50{\times}$ over the
unfused baseline at $L{=}16384$, $d{=}128$, with $129{\times}$ less
high-bandwidth memory (HBM) traffic; the speedup grows linearly with
sequence length.
Code is available upon request.
\end{abstract}

\section{Introduction}

Modern GPU kernel libraries for large language model inference (cuDNN,
hipBLAS, Triton, fused attention~\cite{dao2022flashattention}) achieve high
performance through aggressive autotuning: exhaustive search over tile sizes,
register blocking, and memory layouts.
This approach yields excellent results and continues to improve with
advances in automated search and machine learning compilers.

We complement this landscape by introducing a mathematical description
of tensor operators combined with an abstracted, essential hardware
descriptor: local data share capacity, vector register budget, matrix
fused multiply-add instruction shape, and peak bandwidth.
Together, operator and descriptor define the optimization problem
precisely enough that tile sizes follow analytically, without search.
The result is a \emph{default implementation} that is correct,
predictable, and immediately deployable.

\textbf{Thesis.}
For general matrix multiply (GEMM), 2D convolution, softmax, and fused
attention, all tile sizes are determined by three hardware constraints
applied in order.
The local data share (LDS), the thread-shared memory space, fixes the
reduction tile analytically.
The vector general-purpose registers (VGPRs), the per-thread register
space, fix the parallel tile in closed form from the hardware descriptor
and the matrix fused multiply-add (MFMA) instruction shape.
The number of compute units (CUs) determines occupancy and flags when
tile choices under-utilize the device.
The derivation is compositional and hierarchical: fused attention chains
two GEMM operations with an online softmax, and the VGPR constraint must
account for all live arrays at the synchronization point between the two
GEMMs, the tightest register moment in the kernel.
The tree-based algorithm description makes each constraint explicit,
checkable, and composable across operator levels.

\textbf{Contributions.}
\begin{itemize}
  \item A hierarchical computational model (algorithm tree + visitor)
        that separates the description of an operator from its
        hardware mapping (\S\ref{sec:model}).
  \item A hardware descriptor abstraction (\texttt{MFMAAtom},
        \texttt{HardwareDescriptor}) that encodes the essential
        hardware parameters and enables portability across CDNA GPUs
        (\S\ref{sec:hardware}).
  \item Closed-formula tile derivation for GEMM, 2D convolution,
        softmax, and fused attention, connecting the two above
        (\S\ref{sec:feasibility}--\ref{sec:flash}).
  \item A \textsc{FeasibilityVisitor} that checks VGPR and LDS constraints
        at each tree node and derives \texttt{tile\_Q\_max} and
        \texttt{tile\_KV\_max} parametrically (\S\ref{sec:feasibility}).
  \item Measurements on an AMD CDNA GPU validating the full derivation:
        GEMM at 12.3~TFLOPS (53\% of MFMA peak); fused attention at $1.50{\times}$ over the
        unfused baseline at $L{=}16384$ with $129{\times}$ less
        high-bandwidth memory (HBM) traffic; and a documented correction
        to the matrix fused multiply-add register layout as described in
        the AMD instruction set architecture (ISA) reference
        (\S\ref{sec:mfma_layout}, \S\ref{sec:results}).
\end{itemize}

We are optimistic that the principal contribution of this work extends
beyond the operators and the GPU we studied.
The methodology rests on the ability of human and AI collaboration to
produce precise mathematical descriptions of complex functions and
clean, portable mappings to hardware.
Given that description, the derivation is mechanical and the code
follows.

\section{Computational Model}
\label{sec:model}

A tensor operator is described as a hierarchical algorithm tree.
Each node in the tree represents a level of the computation: the
outermost level maps to the device (all compute units working in
parallel), the next level to a single compute unit and its local
memory, the next to a wavefront and its registers, and the leaf to a
single hardware instruction.
Each level partitions its inputs into tiles, performs a local
computation, and accumulates into an output.
The tree is never fully materialized; it is traversed lazily by a
\emph{visitor} that either executes the computation (for validation)
or emits code (for deployment).

Formally, a node at level $\ell$ carries matrix views $A_\ell$,
$B_\ell$, $C_\ell$ as windows into the backing store, with shapes
derived from the tile policy at that level.
The \textsc{ComputeVisitor} fires a local matrix multiply at each node;
the \textsc{EmitVisitor} generates the corresponding GPU kernel loop.
The two visitors share the same tree, guaranteeing that what is
validated is what is compiled.

When the problem size is not a perfect multiple of the tile size,
the last tile in each dimension is clipped to the valid extent of
the matrix view through \texttt{min\_} and \texttt{max\_} bounds
carried by each node.
The computation is correct by construction: no padding decisions are
made at the algorithm level.
The emitted kernel guards out-of-bounds accesses with a predicate
derived from the clipped extent; the padding value is zero, which is
a no-op for accumulation.
This design separates the tiling policy (how large a tile is) from
the boundary policy (what happens at the edge), and keeps both
visible in the tree.

The tile policy at each level is the subject of this paper.
For a given operator and hardware descriptor, we derive tile shapes
analytically so that: the shared memory at each compute unit is filled
exactly (no wasted capacity); the registers at each thread are not
exceeded (no spilling); and the number of active compute units meets
or exceeds a minimum occupancy threshold.
When these three conditions are satisfied the tree is \emph{feasible},
and the corresponding kernel is correct and deployable without further
search.

\textbf{Example: GEMM tree.}
Figure~\ref{fig:gemm_tree} shows the algorithm tree for
$C = A \cdot B$ with $A{\in}\mathbb{R}^{4096\times 4096}$,
$B{\in}\mathbb{R}^{4096\times 4096}$, derived for our CDNA GPU.
Each level is a tiled partition of the level above; the visitor
fires at the chosen level and stops, leaving the remaining levels to
the GPU runtime (block index, thread index).

\begin{figure}[h]
\centering
\begin{lstlisting}[language={}, frame=single, numbers=none,
                   basicstyle=\ttfamily\footnotesize]
TiledComputation  C = A @ B
  A:(4096,4096)  B:(4096,4096)  C:(4096,4096)
  CU grid: 32x32 (1024 tiles) [parallel -- one BlockGemmNode per CU]
  |
  +-- for I in 32, J in 32 [parallel] / for K in 128 [sequential]
  |       C[I,J] += A[I,K] @ B[K,J]
  |
  +-- BlockGemmNode  [block / CU]
  |     A:(128,32)  B:(32,128)  C:(128,128)
  |     LDS = 32 KB  (A+B tiles; C in registers)
  |     K iterations per CU: 128
  |
  +-- WarpGemmNode  [warp]
  |     A:(32,8)  B:(8,32)  C:(32,32)
  |     K iterations per warp: 4
  |     warp tiles per block: 4x4 = 16
  |
  +-- AtomHint  [leaf -- hardware instruction]
        MFMA/WMMA 16x16x4  dtype=f32
        K iterations per atom: 2
        atom calls per warp:   2x2 = 4
\end{lstlisting}
\caption{Algorithm tree for $4096^3$ GEMM, generated directly from
the code (\texttt{TiledComputation.tree()}).
Tile sizes at each level are derived analytically from the hardware
descriptor.
The \textsc{ComputeVisitor} validates at any chosen level;
the \textsc{EmitVisitor} generates the GPU kernel.}
\label{fig:gemm_tree}
\end{figure}

\section{Hardware Model}
\label{sec:hardware}

For validation, we target an AMD CDNA GPU with the following
characteristics: 64~KB LDS per compute unit (CU), 256 VGPRs per
thread, 64-thread wavefronts, 120 CUs, 1.2~TB/s HBM bandwidth,
11.5~TFLOPS scalar single-precision compute, and 23.1~TFLOPS via
matrix fused multiply-add instructions (two FMAs per clock).

We encode and abstract these hardware parameters in a
\texttt{HardwareDescriptor} and each matrix fused multiply-add
instruction, the basic computational atom on modern GPU
architectures, in an \texttt{MFMAAtom}.
The interface is intentionally minimal: only the quantities that
appear in the tile derivation are present.

\begin{lstlisting}[language=Python]
# Target A: AMD CDNA GPU (our validation platform)
gpu_a = HardwareDescriptor(
  lds_bytes=64*1024,  vgpr_per_thread=256,
  wavefront_size=64,  num_cus=120,
  hbm_bandwidth_tb=1.2, peak_tflops_f32=11.5,
  atoms=[MFMAAtom(m=16, n=16, k=4,
                  dtype_in='f32', dtype_acc='f32')],
)

# Target B: a second CDNA GPU (larger, more CUs)
gpu_b = HardwareDescriptor(
  lds_bytes=64*1024,  vgpr_per_thread=256,
  wavefront_size=64,  num_cus=304,
  hbm_bandwidth_tb=5.3, peak_tflops_f32=383.0,
  atoms=[MFMAAtom(m=16, n=16, k=4,
                  dtype_in='f32', dtype_acc='f32')],
)

# Same derivation, different descriptor -- tiles adapt automatically
tiles_a = derive_tiles(gpu_a, L=16384, d=128)
tiles_b = derive_tiles(gpu_b, L=16384, d=128)
# tiles_a: tile_Q=32, tile_KV=64  (120 CUs, 0.27/CU -> low occupancy flagged)
# tiles_b: tile_Q=32, tile_KV=64  (304 CUs, 1.69/CU -> good occupancy)
# Tile sizes identical: both are LDS- and VGPR-limited, not CU-limited.
# Only the occupancy assessment changes.
\end{lstlisting}

The derivation function takes a descriptor and problem dimensions;
it returns tile sizes, feasibility status, and the binding constraint.
No other code changes when moving between targets.
For \texttt{mfma\_f32\_16x16x4f32}: each thread holds
$16{\times}4/64 = 1$ A-fragment float,
$4{\times}16/64 = 1$ B-fragment float, and
$16{\times}16/64 = 4$ C/D-fragment floats.

\subsection{A Pitfall for the Code Generator}
\label{sec:mfma_layout}

The \texttt{MFMAAtom} encodes the instruction shape $(m, n, k)$, but
shape alone is not sufficient for a correct code generator.
The generator must also know how output elements are distributed
across threads. This distribution is not derivable from the
shape parameters alone.

A natural assumption is that thread \texttt{tid} owns output row
$\texttt{tid} \bmod m$ and output column group $\texttt{tid} \div m$.
On our CDNA GPU, \texttt{mfma\_f32\_16x16x4f32} does the opposite:
\begin{equation}
  \texttt{acc}[f] = C[\texttt{col\_group}{\times}4 + f][\texttt{row\_in\_m}]
  \quad
  \text{where } \texttt{col\_group} = \lfloor \texttt{tid}/16 \rfloor,\;
  \texttt{row\_in\_m} = \texttt{tid} \bmod 16
\end{equation}
That is, $\texttt{col\_group}$ drives output \emph{rows} and
$\texttt{row\_in\_m}$ drives output \emph{columns}, the transpose
of the naive expectation.
We confirmed this with the structured test $Q[i,:]{=}i$, $K{=}I$
$\Rightarrow$ $C[i,:]{=}i$: if the layout were as documented,
the output would be the identity; the actual output revealed the
transposition.

The consequence for a code generator is concrete.
When two matrix multiply instructions are chained, with the output of the
first feeding the input of the second as in fused attention, the
generator must either transpose the intermediate result through shared
memory, or derive the A-fragment index of the second instruction from
the confirmed C-fragment layout of the first, not from the
nominal shape.
A generator that relies on the nominal shape alone will produce a
kernel that compiles, runs, and produces wrong results silently, a
failure mode that is difficult to diagnose without a structured test
like the one above.
The general principle is that any code generator targeting tensor
instructions must treat fragment ownership as an empirical property
of the hardware, not as a consequence of the instruction name or
shape parameters.
Encoding it explicitly in the hardware descriptor, as we do in
\texttt{MFMAAtom}, makes the assumption visible, testable, and
portable to any future target where the layout may differ again.

\section{Tile Derivation}
\label{sec:feasibility}

Given the algorithm tree (\S\ref{sec:model}) and the hardware
descriptor (\S\ref{sec:hardware}), tile sizes at each level follow
from three constraints applied in a fixed order.
The process is neither top-down nor bottom-up: it is \emph{CU-centric}.

The compute unit is the natural anchor because it owns the two binding
resources: the local data share and the vector registers.
The derivation starts at the CU, satisfies its constraints to obtain
tile sizes, then propagates outward in both directions: upward to
check that enough tiles exist to cover the problem with reasonable
occupancy, and downward to verify alignment with the atom instruction
shape.
The runtime traverses the tree top-down; the tile derivation builds
it CU-first.

We illustrate the derivation using the $4096^3$ GEMM of
Figure~\ref{fig:gemm_tree}, where the tree has three levels:
device (the full grid), block (one CU), and warp (one wavefront),
with the atom as the leaf.

\textbf{LDS constraint $\Rightarrow$ CU tile.}
The \texttt{BlockGemmNode} (the CU level) loads tiles of $A$ and $B$
into LDS simultaneously.
The CU tile dimensions $(t_M, t_N, t_K)$ must satisfy:
$(t_M + t_N) \times t_K \times \texttt{elem} \leq \text{LDS}$.
Setting this equal to the budget and choosing $t_M = t_N$:
\begin{equation}
  t_M = t_N = \left\lfloor \sqrt{\frac{\text{LDS}/2}{\texttt{elem}}} \right\rfloor,
  \quad
  t_K = \left\lfloor \frac{\text{LDS}/2}{(t_M + t_N)\cdot\texttt{elem}} \right\rfloor
  \label{eq:tile_cu}
\end{equation}
For our CDNA GPU (64~KB LDS, \texttt{elem}~$=4$~bytes):
$t_M = t_N = 128$, $t_K = 32$, exactly the \texttt{BlockGemmNode}
tile in Figure~\ref{fig:gemm_tree}.

\textbf{VGPR constraint $\Rightarrow$ warp tile.}
The \texttt{WarpGemmNode} (the wavefront level) accumulates a sub-tile
of $C$ in per-thread registers throughout the $K$ loop.
The warp tile $(t_{wm}, t_{wn}, t_{wk})$ must fit within the VGPR
budget: each thread holds $(t_{wm} \times t_{wn}) / \text{wf}$
accumulator floats, where wf~$=64$ is the wavefront size.
For our warp tile $(32, 32, 8)$: $32\times32/64 = 16$ floats per
thread, well within the 256-VGPR budget.
With MFMA dimensions $(m, n, c) = (16, 16, 4)$ and $u = t_{wm}/m$:
\begin{equation}
  V(u) = u \cdot \frac{t_{wn}}{n} \cdot c + V_{\text{overhead}} \leq V_{\text{budget}}
  \label{eq:vgpr}
\end{equation}
The warp tile is the largest $(t_{wm}, t_{wn})$ satisfying this
inequality, aligned to the atom dimensions $(m, n)$.

\textbf{Occupancy constraint: CU count.}
The grid contains $\lceil M/t_M \rceil \times \lceil N/t_N \rceil$
blocks. For $M=N=4096$ and $t_M=t_N=128$: $32\times32 = 1024$ blocks
for 120 CUs, giving 8.5 blocks per CU, good occupancy.
If this ratio falls below 1, the derivation reduces tile sizes until
occupancy is acceptable.

\textbf{Atom alignment.}
Finally, tile dimensions must be multiples of the atom shape $(m, n, k)
= (16, 16, 4)$: $t_M=128$, $t_N=128$, $t_K=32$ are all divisible.
No padding is needed; every atom fires on valid data.

The \textsc{FeasibilityVisitor} automates these four checks for any
operator and hardware descriptor, returning the tile sizes or the
binding constraint that prevents feasibility.
For fused attention, the VGPR constraint becomes significantly tighter
because two GEMMs are chained and multiple intermediate arrays are
simultaneously live in registers at the synchronization point between
them; we present the full derivation in \S\ref{sec:flash}.

\section{Operators}
\label{sec:operators}

We present four operators in order of increasing complexity.
Each is described as an algorithm tree; the tile policy and hardware
constraints are derived analytically; and the resulting kernel is validated
against a double-precision reference on our CDNA GPU.

\subsection{GEMM}

For $C = A \cdot B$ with $A{\in}\mathbb{R}^{M\times K}$,
$B{\in}\mathbb{R}^{K\times N}$, the CU and warp tile sizes follow
from the derivation of \S\ref{sec:feasibility}, with Figure~\ref{fig:gemm_tree}
showing the result for $M=N=K=4096$.

For code generation we use FlyDSL~\cite{flydsl}, a Python-embedded
domain-specific language that traverses the algorithm tree and emits
GPU kernels.
FlyDSL illustrates a key property of the computational model: because
the operator is described as a structured, machine-readable tree, the
code generation step is a traversal, not a translation.
Any backend that can consume the tree (FlyDSL, a Triton emitter, a
vendor library wrapper, or an agentic search over implementation
variants) receives the same well-defined specification.
The tree both anchors the closed-formula default and defines the
solution space for any subsequent automated exploration.

Result: $4096^3$ GEMM at \textbf{12.3~TFLOPS}, validated at machine
precision vs double-precision reference.
Table~\ref{tab:gemm_perf} summarizes the result against the relevant
hardware peaks.

\begin{table}[h]
\centering
\small
\caption{GEMM performance on our CDNA GPU.
MFMA FP32 peak = 23.1~TFLOPS (matrix fused multiply-add, two FMAs per clock);
scalar FP32 peak = 11.5~TFLOPS (reference only -- not achievable by MFMA kernels).
At 53\% of MFMA peak on a first implementation, the remaining gap is
explained by incomplete pipeline utilization: issuing multiple independent
MFMA instructions per wavefront hides latency and closes the gap.}
\label{tab:gemm_perf}
\begin{tabular}{@{}lrr@{}}
\toprule
 & TFLOPS & \% MFMA peak \\
\midrule
$4096^3$ GEMM (measured) & 12.3 & 53\% \\
MFMA FP32 peak           & 23.1 & 100\% \\
\bottomrule
\end{tabular}
\end{table}

\subsection{2D Convolution}

A 2D convolution $Y = T(X, W)$ with input
$X \in \mathbb{R}^{N \times IC \times IH \times IW}$
(batch $N$, input channels $IC$, spatial height $IH$, width $IW$)
and filter $W \in \mathbb{R}^{OC \times IC \times KH \times KW}$
(output channels $OC$, kernel height $KH$, width $KW$) admits
several implementation strategies: direct convolution, lowering to GEMM
via im2col, Winograd, and frequency-domain methods.
Each trades arithmetic against memory traffic differently, and each
maps to a different algorithm tree.
We focus on the direct implementation, applying each filter directly
to its corresponding input patch without data rearrangement, because
it is the most general, requires no preprocessing of inputs or weights,
and exposes the tiling structure most clearly.
The filter pre-packing strategy we describe below is the one degree of
freedom available at deployment time: the filter $W$ is constant during
inference and can be stored in any layout that benefits the computation.

\textbf{Halo analysis.} Each output tile $(t_{OH}, t_{OW})$ requires an
input halo of $(t_{OH} + (KH{-}1), t_{OW} + (KW{-}1))$.
The reuse factor for a $3{\times}3$ filter is:
\begin{equation}
  R = \left(1 + \frac{KH-1}{t_{OH}}\right)\!\left(1 + \frac{KW-1}{t_{OW}}\right)
\end{equation}
Larger tiles reduce re-reads of the halo at the cost of more LDS.

\textbf{Filter pre-packing.} For inference, the filter $W$ is constant
and can be stored in any layout at deployment time.
We use the layout:
\[
  (OC/t_{OC},\; IC/t_{IC},\; KH,\; KW,\; t_{OC},\; t_{IC})
\]
which groups the output- and input-channel tile elements contiguously,
giving each CU a single contiguous read for its assigned tile.

\textbf{Spatial-to-channel replication.} For small input-channel inputs
(e.g.,\ $IC{=}3$ for RGB), we apply the pixel-shuffle
trick~\cite{shi2016}: replicate spatially to $IC{=}12$, reducing partial
sums from $KH{\times}KW$ to 1 while maintaining correctness.

Result: tiled $3{\times}3$ kernel at \textbf{1.7$\times$} over naive
baseline on our CDNA GPU; all 42 algorithm family tests pass.

\subsection{Online Softmax}

Softmax is always bandwidth-bound (0.2~FLOPs/byte).
The online algorithm~\cite{milakov2018} accumulates a running maximum
$m_{\text{run}}$ and denominator $d_{\text{run}}$ in registers, applying
a correction factor $s = \exp(m_{\text{run}} - m_{\text{new}})$ when a
new tile maximum exceeds the previous running value.

The merge operator $\oplus$ is associative:
\begin{align}
  (m_a, d_a) \oplus (m_b, d_b) &= \bigl(m_{\text{new}},\;
    d_a \exp(m_a{-}m_{\text{new}}) + d_b \exp(m_b{-}m_{\text{new}})\bigr)
\end{align}
where $m_{\text{new}} = \max(m_a, m_b)$.
Associativity enables split-N parallelism via a binary reduction.

A practical note on boundary padding: out-of-bounds tile elements must be
padded with $-\infty$ (approximated as $-10^{38}$ in single precision), not
with zero.
Zero padding produces $\exp(0 - m) > 0$, which contributes a spurious
positive weight to the denominator and corrupts the output.
Hardware architectures that pad with zeros require an explicit mask applied
before or after the exponential.

Validated at machine precision on our CDNA GPU at $512{\times}512$.

\section{Fused Attention}
\label{sec:flash}

Multi-head attention (MHA) is the core operation of the transformer
architecture~\cite{dao2022flashattention}.
Each head computes a weighted sum of value vectors, where the weights
are determined by the similarity between query and key vectors.
Formally, given queries $Q \in \mathbb{R}^{L \times d}$,
keys $K \in \mathbb{R}^{L \times d}$, and
values $V \in \mathbb{R}^{L \times d}$, all of sequence length $L$
and head dimension $d$, and the output is:
\begin{equation}
  R = \text{softmax}\!\left(\frac{QK^T}{\sqrt{d}}\right) V
  \quad R \in \mathbb{R}^{L \times d}
\end{equation}
The intermediate attention score matrix $QK^T \in \mathbb{R}^{L\times L}$
is the computational bottleneck: materializing it to HBM requires
$O(L^2)$ memory and traffic that grows quadratically with sequence
length.

Fused attention~\cite{dao2022flashattention} eliminates this
bottleneck by streaming $K$ and $V$ in tiles through a single loop,
keeping the score tile $S_t = Q_{\text{tile}} K_t^T / \sqrt{d}$
register-resident at all times and never writing it to HBM.
The savings ratio relative to the unfused computation is:
\begin{equation}
  \text{savings} = \frac{L^2 \cdot \text{elem}}{4 \cdot L \cdot d \cdot \text{elem}} = \frac{L}{4d}
  \label{eq:savings}
\end{equation}
This grows linearly with $L$ and is independent of tile sizes. It is
a structural property of the algorithm, not of the implementation.

\textbf{Block structure.}
Partition $Q$, $K$, $V$ into tiles $Q_i \in \mathbb{R}^{t_Q \times d}$,
$K_j, V_j \in \mathbb{R}^{t_{KV} \times d}$.
The full score matrix has block structure:
\[
  \begin{pmatrix} Q_0 \\ Q_1 \\ \vdots \\ Q_n \end{pmatrix}
  \begin{pmatrix} K_0 & K_1 & \cdots & K_n \end{pmatrix}
  =
  \begin{pmatrix}
    Q_0 K_0 & Q_0 K_1 & \cdots & Q_0 K_n \\
    Q_1 K_0 & Q_1 K_1 & \cdots & Q_1 K_n \\
    \vdots  &          & \ddots & \vdots  \\
    Q_n K_0 & Q_n K_1 & \cdots & Q_n K_n \\
  \end{pmatrix}
\]
Each output tile $R_i$ depends only on row $i$ of this matrix and the
corresponding rows of $V$.
For $R_0$:
\begin{equation}
  R_0 = \frac{
    \sum_j \exp(Q_0 K_j)\, V_j
  }{
    \sum_j \exp(Q_0 K_j)
  }
  \label{eq:r0}
\end{equation}
where the exponential and the sum in the denominator are row-wise.
The key observation is that Eq.~\eqref{eq:r0} is a sum over $j$:
each term $\exp(Q_0 K_j) V_j$ can be accumulated independently,
and the denominator $\sum_j \exp(Q_0 K_j)$ is a running scalar.
This is the associativity that allows streaming $K$ and $V$ tile by
tile while $Q_0$ remains in registers, without ever materializing the
full $L \times L$ score matrix.

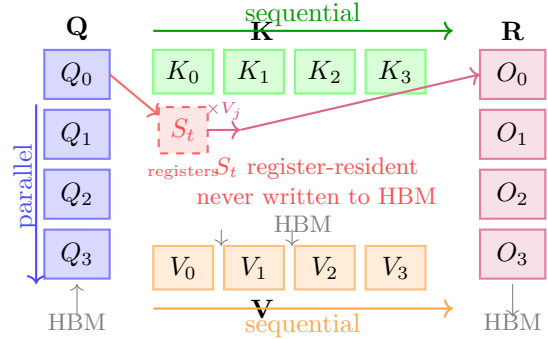
\begin{figure}[h]
\centering
\begin{tikzpicture}[scale=0.72, every node/.style={font=\small}]

  \foreach \i in {0,1,2,3} {
    \pgfmathsetmacro{\y}{-\i*1.1}
    \draw[fill=blue!15, draw=blue!50, thick]
      (0, \y) rectangle (1.2, \y+0.9);
    \node at (0.6, \y+0.45) {$Q_\i$};
  }
  \node[blue!70, rotate=90] at (-0.35, -1.6) {\small parallel};
  \draw[blue!70, ->, thick] (-0.15, -0.1) -- (-0.15, -3.4);
  \node[above] at (0.6, 0.9) {\textbf{Q}};

  \foreach \j in {0,1,2,3} {
    \pgfmathsetmacro{\x}{2.0 + \j*1.3}
    \draw[fill=green!15, draw=green!50, thick]
      (\x, 0.1) rectangle (\x+1.1, 0.9);
    \node at (\x+0.55, 0.5) {$K_\j$};
  }
  \node[above] at (4.0, 0.9) {\textbf{K}};
  \draw[green!60!black, ->, thick] (2.0, 1.25) -- (7.5, 1.25);
  \node[green!60!black] at (4.75, 1.55) {\small sequential};

  \foreach \j in {0,1,2,3} {
    \pgfmathsetmacro{\x}{2.0 + \j*1.3}
    \draw[fill=orange!15, draw=orange!50, thick]
      (\x, -3.5) rectangle (\x+1.1, -2.7);
    \node at (\x+0.55, -3.1) {$V_\j$};
  }
  \node[below] at (4.0, -3.5) {\textbf{V}};
  \draw[orange!70, ->, thick] (2.0, -3.85) -- (7.5, -3.85);
  \node[orange!70] at (4.75, -4.15) {\small sequential};

  \draw[fill=red!10, draw=red!60, thick, dashed]
    (2.1, -0.15) rectangle (3.0, -1.0);
  \node[red!70] at (2.55, -0.57) {$S_t$};
  \node[red!70, font=\tiny] at (2.55, -1.35) {registers};

  \foreach \i in {0,1,2,3} {
    \pgfmathsetmacro{\y}{-\i*1.1}
    \draw[fill=purple!12, draw=purple!50, thick]
      (8.0, \y) rectangle (9.2, \y+0.9);
    \node at (8.6, \y+0.45) {$O_\i$};
  }
  \node[above] at (8.6, 0.9) {\textbf{R}};

  \draw[->, thick, red!60] (1.2, 0.45) -- (2.1, -0.3);

  \draw[->, thick, purple!60] (3.0, -0.57) -- (3.6, -0.57)
    node[midway, above, font=\tiny, purple!70] {$\times V_j$};
  \draw[->, thick, purple!60] (3.6, -0.57) -- (8.0, 0.45);

  \node[red!70, font=\footnotesize, align=center] at (5.0, -1.5)
    {$S_t$ register-resident\\never written to HBM};

  \node[font=\footnotesize, gray] at (0.6, -4.1) {HBM};
  \draw[gray, ->] (0.6, -4.0) -- (0.6, -3.4);
  \node[font=\footnotesize, gray] at (4.75, -2.3) {HBM};
  \draw[gray, ->] (3.25, -2.4) -- (3.25, -2.7);
  \draw[gray, ->] (4.55, -2.4) -- (4.55, -2.7);
  \node[font=\footnotesize, gray] at (8.6, -4.1) {HBM};
  \draw[gray, ->] (8.6, -3.4) -- (8.6, -4.0);

\end{tikzpicture}
\caption{Tile structure of fused attention.
Q tiles (blue) are processed in parallel, one per CU.
Within each Q tile, K and V tiles (green, orange) stream sequentially
from high-bandwidth memory (HBM).
The score tile $S_t = Q_i K_j^T / \sqrt{d}$ lives entirely in
registers and is never written to HBM.
The output accumulator $O_i$ is updated in-place and written once
at the end.
HBM traffic is $O(L \cdot d)$, not $O(L^2)$.}
\label{fig:fused_attn}
\end{figure}

\begin{algorithm}
\caption{Fused attention (one Q tile). $d$: head dimension. $m$: running row maximum. $\delta$: running softmax denominator.}
\begin{algorithmic}[1]
\STATE $m \leftarrow -\infty$, $\delta \leftarrow 0$, $O \leftarrow 0$
\FOR{each key/value tile $t$}
  \STATE $S_t \leftarrow Q_{\text{tile}} K_t^T / \sqrt{d}$ \hfill \textit{GEMM, registers}
  \STATE $m_{\text{new}} \leftarrow \max(m, \text{rowmax}(S_t))$ \hfill \textit{warp reduce}
  \STATE $s \leftarrow \exp(m - m_{\text{new}})$ \hfill \textit{correction factor}
  \STATE $E_t \leftarrow \exp(S_t - m_{\text{new}})$ \hfill \textit{in-place}
  \STATE $\delta \leftarrow \delta \cdot s + \text{rowsum}(E_t)$
  \STATE $O \leftarrow O \cdot s + E_t V_t$ \hfill \textit{GEMM, registers}
  \STATE $m \leftarrow m_{\text{new}}$
\ENDFOR
\STATE \textbf{return} $O / \delta$
\end{algorithmic}
\end{algorithm}

\subsection{MFMA Kernel}

We implement two kernels.
The \textbf{MFMA+FMA kernel} uses matrix fused multiply-add (MFMA)
for the first GEMM ($Q \cdot K^T$) and scalar fused multiply-add
(FMA) operations for the second ($E_t \cdot V$).
After the warp reduce, $E_t$ is written to LDS row-major and the
second GEMM is computed element-wise per thread.

The \textbf{MFMA+MFMA fused kernel} uses MFMA for both GEMMs.
$E_t$ is written to LDS \emph{transposed}, stored at
\texttt{e\_lds[col $\times$ TILE\_Q + row]}, so the second GEMM
reads it as A-fragments with the same index formula as the first GEMM
reads $Q$:
\begin{equation}
  a = \texttt{e\_lds}[\texttt{row\_in\_m} \times t_Q + k_s \cdot m + \texttt{col\_group}]
\end{equation}
No scalar loops remain.
Both kernels are validated at machine precision; identical results
confirm the transposed layout is correct.
A complete analysis of precision and numerical behavior for attention
computations, including the effect of accumulator precision on
key-value cache compression, is provided in~\cite{dalberto2026ablation}.

\subsection{Results}
\label{sec:results}

Table~\ref{tab:tiles} shows the tile sizes derived by the
\textsc{FeasibilityVisitor} for two representative head dimensions,
illustrating how the VGPR constraint tightens as $d$ grows.

\begin{table}[h]
\centering
\caption{Closed-formula tile sizes for fused attention on our CDNA GPU.
VGPRs measured at the synchronization point between the two GEMMs,
where $Q_{\text{reg}}$, $S_t$, and $O_{\text{acc}}$ are all live.}
\label{tab:tiles}
\begin{tabular}{@{}rrrrc@{}}
\toprule
$d$ & $t_{\text{KV}}$ & $t_{Q,\max}$ & VGPRs & Feasible \\
\midrule
64  & 128 & 48 & 214/256 & \checkmark \\
64  & 128 & 64 & 280/256 & $\times$ \\
128 & 64  & 32 & 180/256 & \checkmark \\
128 & 64  & 48 & 262/256 & $\times$ \\
\bottomrule
\end{tabular}
\end{table}

Table~\ref{tab:sweep} shows the sweep over $L$ for $d{=}128$ on our CDNA GPU.
The unfused baseline uses three separate tiled GEMM kernels with the
$L{\times}L$ attention matrix written to HBM between steps.

\begin{table}[h]
\centering
\small
\caption{Fused attention (MFMA+MFMA) vs unfused. $d{=}128$, $t_Q{=}32$,
$t_{\text{KV}}{=}64$, our CDNA GPU.
Speedup = unfused / fused.}
\label{tab:sweep}
\begin{tabular}{@{}rrrr@{}}
\toprule
$L$ & Fused (ms) & Unfused (ms) & Speedup \\
\midrule
4{,}096  &  27.9 &  26.6 & 0.95$\times$ \\
8{,}192  &  83.5 & 104.6 & 1.25$\times$ \\
16{,}384 & 279.3 & 418.1 & 1.50$\times$ \\
32{,}768 & 991.1 & 1{,}666.6 & 1.68$\times$ \\
\bottomrule
\end{tabular}
\end{table}

\textbf{Crossover at $L{\approx}4096$.}
Below the crossover, the LDS overhead of writing $E_t^T$ before the second
GEMM exceeds the HBM savings from avoiding the attention matrix.
Above it, HBM savings dominate and speedup grows with $L$.

\textbf{Trend.}
Speedup grows from $1.25{\times}$ at $L{=}8192$ to $1.68{\times}$ at
$L{=}32768$, consistent with the $L/(4d)$ savings formula
(Eq.~\eqref{eq:savings}).
At $L{=}32768$, $d{=}128$, fused attention avoids writing and reading a
4~GB attention matrix, traffic that would not fit in a single device's HBM
for batched inference.

\textbf{Remaining gap.}
The MFMA+MFMA fused kernel achieves $1.50{\times}$ over unfused at $L{=}16384$.
This is short of the $\sim$33$\times$ theoretically available at the roofline.
The bottleneck is LDS bandwidth: writing $E_t^T$ (24~KB for $t_Q{=}32$,
$t_{\text{KV}}{=}64$) and loading $V$ each key/value step adds
$\sim$10~KB/step of LDS traffic absent from the roofline model.
Eliminating this roundtrip by fusing the softmax update and the second
GEMM at the register level is left as future work.

\textbf{HBM traffic.}
At $L{=}16384$, fused attention reads/writes 34~MB (Q, K, V, R once each).
Unfused reads/writes 4{,}329~MB ($129{\times}$ more), dominated by the
attention matrix $S$ at 1{,}074~MB.
At $L{=}32768$, unfused traffic reaches 17~GB for a single head.

\section{Related Work}

\textbf{Fused attention.}
Dao et al.~\cite{dao2022flashattention} introduced the fused online
softmax approach; a follow-up~\cite{dao2023flashattention2}
added GEMM scheduling improvements.
Our contribution is the analytical tile derivation and the MFMA layout
documentation, not a new algorithm.

\textbf{Autotuning.}
Triton~\cite{tillet2019triton}, TVM~\cite{chen2018tvm}, and
Halide~\cite{ragan2013halide} search over tile configurations.
Our closed-formula defaults provide a sound starting point for such
searches, reducing the space to explore and guaranteeing a correct
baseline at zero search cost.

\textbf{Analytical tile sizing.}
ATLAS~\cite{whaley2001atlas} and Spiral~\cite{puschel2005spiral}
pioneered algorithm-by-hardware co-design.
Our contribution is extending this to fused operators with explicit
register constraint propagation through the algorithm tree.

\textbf{Hand-optimized AMD GPU kernels.}
HipKittens~\cite{hu2025hipkittens} provides C++ primitives for high-performance
GEMM and attention kernels on AMD CDNA GPUs, including asynchronous load/store
patterns and multi-wave ping-pong scheduling to saturate the MFMA pipeline.
These techniques address exactly the gap between our 53\% MFMA utilization
and the theoretical peak: issuing multiple independent MFMA instructions
per wavefront hides instruction latency.
Our closed-formula defaults could serve as the tile-size starting point for
such hand-optimized implementations, separating the derivation of correct
tile sizes from the engineering of the instruction schedule.

\textbf{Optimized BLAS libraries.}
OpenBLAS~\cite{openblas} provides highly tuned GEMM kernels for CPU targets,
using SIMD fused multiply-add (FMA) instructions and analytically derived
blocking parameters.
Our work extends a similar philosophy, deriving tile sizes from hardware
parameters rather than searching, to GPU tensor instructions and fused
operators such as attention, where the register layout constraints are
qualitatively more complex.
The C-fragment layout discrepancy we document for chained MFMA GEMMs has
no direct analogue in CPU SIMD work and is specific to GPU tensor instruction
architectures.

\section{Conclusion}

Tile sizes for GEMM, 2D convolution, softmax, and fused attention
follow from a CU-centric derivation: the local data share determines
the CU tile, the vector registers determine the warp tile, and the
number of compute units bounds the minimum problem granularity.
The \textsc{FeasibilityVisitor} automates this derivation: swap the
hardware descriptor and the tiles update without any re-derivation.

Fused attention with MFMA achieves $1.50{\times}$ over unfused
attention at $L{=}16384$, $d{=}128$ on our CDNA GPU, with $129{\times}$ less
HBM traffic; speedup reaches $1.68{\times}$ at $L{=}32768$.
A practical finding accompanies the performance results: the MFMA
register layout on our CDNA GPU differs from the AMD ISA documentation;
we provide a minimal reproducible test and the correct formula.

We are optimistic that the principal contribution of this work extends
beyond the operators and the GPU we studied.
The methodology rests on the ability of human and AI collaboration to
produce precise mathematical descriptions of complex functions and
clean, portable mappings to hardware.
Given that description, the derivation is mechanical and the code
follows.
We expect the approach to generalize to any operator expressible as a
hierarchical computation tree, on any architecture for which an
essential hardware descriptor can be written.

Future work: eliminate the $E_t^T$ LDS roundtrip to close the gap
toward the roofline; extend to multi-head batching, further operators,
and additional GPU targets.

\bibliographystyle{plain}
\bibliography{refs}

\end{document}